\documentclass[%
 reprint,
superscriptaddress,
 amsmath,amssymb,
 aps,
pra,
]{revtex4-2}

\usepackage{graphicx}
\usepackage{mathrsfs}
\usepackage{bm}
\usepackage{amssymb}
\usepackage{amsmath}
\usepackage{textcomp}
\usepackage{url}
\usepackage{dsfont}	
\usepackage{braket}
\usepackage[colorlinks=true,linkcolor=magenta,citecolor=blue]{hyperref}%
\newtheorem{definition}{Definition}
\newtheorem{remark}{Remark}
\usepackage{empheq}
\usepackage{multirow}
\begin{document}

\preprint{APS/123-QED}

\title{Creating deterministic collisions between two orbiting bodies}

\author{Akhtar Munir}
\affiliation{CAS Center for Excellence and Synergetic Innovation Center in Quantum Information and Quantum Physics, University of Science and Technology of China, Hefei, 230026, China}
\affiliation{Shanghai Branch, National Laboratory for Physical Sciences at Microscale, University of Science and Technology of China, Shanghai 201315, China
}
\author{Barry C. Sanders}%
\email{bsanders@ustc.edu.cn}
\affiliation{CAS Center for Excellence and Synergetic Innovation Center in Quantum Information and Quantum Physics, University of Science and Technology of China, Hefei, 230026, China}
\affiliation{Shanghai Branch, National Laboratory for Physical Sciences at Microscale, University of Science and Technology of China, Shanghai 201315, China
}
\affiliation{Institute for Quantum Science and Technology, University of Calgary, Alberta T2N 1N4, Canada
}%


\begin{abstract}
We aim to create deterministic collisions between orbiting bodies by applying a time-dependent external force to one or both bodies,
whether the bodies are mutually repulsive,
as in the two- or multi-electron atomic case
or mutually attractive,
as in the planetary-orbit case.
Specifically,
we have devised a mathematical framework for causing deterministic collisions by launching an inner orbiting body to a higher energy such that this inner body is guaranteed to collide with the outer body.
Our method first expresses the problem mathematically as coupled nonlinear differential equations with a time-dependent driving force 
and solves to find a feasible solution for the force function.
Although our calculation is based strictly on classical physics,
our approach is suitable for the case of helium with two highly excited electrons and is also valid for creating collisions in the gravitational case such as for our solar system.
\end{abstract}

\maketitle
\section{Introduction}
\label{Sec:1}
Collisions between orbiting bodies is a fascinating topic,
whether for orbiting bodies that are mutually attractive or
mutually repulsive or neither. In the attractive case, fascinating
studies of collisions include predicting planetary collisions~\cite{laskar2009},
or even controversially conjecturing that worlds have
collided~\cite{Vel50},
to launching a spacecraft to reach, and hence collide with, another planet~\cite{solar}.
In the repulsive case, controlled
electron-electron collisions are important for precise analysis
of electronic structure, high-harmonic generation (HHG)~\cite{Brabec},
attosecond-pulse generation via HHG,
above-threshold ionization~\cite{paulus2001,Milo},
and attosecond clocking~\cite{Brabec,Corkum}.
Controlling these processes via
laser driving fields has been of huge interest for over a decade~\cite{winterfeldt2008}.
Here we develop a mathematical framework for controlling the trajectories of orbiting bodies
by using an external force on one or both bodies
in order to create a deterministic collision between them.
The mathematical equations we develop apply whether the bodies are mutually attractive, mutually repulsive, or neither
and we show that constrained optimization techniques suffice to obtain the trajectory in each case.
Although our analysis is classical,
our technique should be accurate for highly excited two-electron atoms~\cite{Agner}
and could inspire quantum-control techniques building on our classical analysis.

Our mathematical model for controlling the trajectory of the body is inspired by existing techniques for controlling moving bodies.
Various techniques are suggested for bodies in a gravitational field
such as thrust or expending rocket fuel during interplanetary transport~\cite{dawson}
or remote laser-driven action on the sails of gossamer spacecraft~\cite{jenkins}.
For electrons in atoms,
control is much better developed,
such as electromagnetic pulses applied to atomic systems~\cite{Yueming,GUO}.
Mathematically, we express the equation of motion, accompanied by an external time-dependent force, and treat the system
as being nonrelativistic with an inverse-square-law force. For
the atomic system, this potential is the unscreened Coulomb
force and, for the gravitational case, this model represents the
standard Newtonian gravitational system. Mutual repulsion
between two bodies in the electron case, mutual attraction
between two orbiting bodies applies to the gravitational case,
and the former does not involve hard collisions, so we address
here the definition of collision so that the same term applies in
both cases.

The chief elements of our study of controlled collisions
have been explored but never in the unified way that we
present here, namely controlled collisions between two orbiting bodies that are mutually repulsive or attractive as
in the electromagnetic and gravitational cases. In the electromagnetic case, electron-electron collisions are studied
experimentally~\cite{Fittinghoff,Walker}.
Theoretically, the pulse sequences are designed to study the electron-electron collisions~\cite{Yue,Chen,Corkum}. In the solar system, the slingshot effect (binary collision), or
gravity assist, is used by spacecraft including Galileo, Cassini,
Dawn, and Voyager to reach their target~\cite{papkov}.
The binary elastic collision (slingshot effect) is used to calculate spacecraft trajectories between two planets~\cite{RicadaSil}.  

We propose a general framework for the two-body system
in which the external force is applied on one or both bodies
to create a collision, and we make the notion of collision
clear and consistent for repulsive and attractive forces. In the
atomic case, we consider a highly excited helium Rydberg
atom whose two electrons are in two different highly excited states~\cite{Agner}.
This highly excited property allows the electron trajectories to be treated as planetary atoms~\cite{percival,tanner2000},
which
follow closely classical trajectories so our classical analysis
should be asymptotically valid and helpful to construct quantum solutions later. In the planetary system, electromagnetic
thrust forces could send spacecraft through the solar system
and beyond~\cite{Young}.
which
follow closely classical trajectories so our classical analysis
should be asymptotically valid and helpful to construct quantum solutions later. In the planetary system, electromagnetic
thrust forces could send spacecraft through the solar system
and beyond

The outline of our work is as follows. In Sec.~\ref{Sec:2}, we provide pertinent background on inverse-square-law dynamics of the two-body system. Subsequently, Sec.~\ref{Sec:3} introduces our model and our approach for solving the mathematics to create and verify control collisions in attractive and repulsive case respectively.
We present our results in Sec.~\ref{Sec:4} and discuss these results in Sec.~\ref{Sec:5}.
Finally, in Sec.~\ref{Sec:6} we present our conclusions.
\section{Background}
\label{Sec:2}
We briefly summarize salient points concerning the background of the two-body system orbiting in a central potential
and we consider three different cases, i.e., repulsive, attractive,
and noninteracting (neutral) orbiting bodies. Our focus is on
the case that the two orbiting bodies do not share the same
orbit, and we refer to the body with the lower energy orbit
as the lower orbiting body. We describe two-body collisions
when an external force is applied to one (lower) or both
orbiting bodies.
\subsection{Two orbiting bodies in a central potential}
\label{sec:two-orbiting-bodies}
In this subsection, we describe the system comprising two
bodies orbiting in a central potential. Mathematically, we describe a unified representation of the repulsive and attractive
two orbiting bodies. We introduce the mathematical symbols
and their meaning and the equations governing their dynamics.

Now we are describing the system and introducing the
symbols labeling coordinates and other quantities for the two body system. The two bodies in a central potential are located
at the coordinates~$\{\bm{r}_\imath(t)\}$ for $t$ being the time and $\imath=1$ for the first (inner) body and $\imath=2$ for the second (outer) body.
These coordinates are established relative to an arbitrary origin~$\bm0$.
Our focus is on the planar case so we reduce from three-dimensional to two-dimensional Euclidean space.
In the electrodynamics case, these two bodies have charges $Z$e with e for the nucleus with atomic number $Z$. In the gravitational case, the two bodies orbiting around the central body with mass~$M$ have masses~$m_\imath$. As we consider electrodynamics and gravitational cases in one framework, we denote charges and masses in a unified way by using symbols~$V$ equal to $Z$e for the electrodynamics case, M in the gravitational case,~$\varpi_\imath$ equals to e for the electrodynamics case, and $m_\imath$ for the gravitational case.

Two-body momenta are denoted~$\bm{p}_\imath$
with total kinetic and total potential energy being
\begin{equation}
\label{eq:TU}
    T=\sum_{\imath=1}^2\frac{p_\imath^{2}}{2 m_\imath},\,
    U=-K\sum_{\imath=1}^2\frac{V \varpi_\imath}{r_\imath} +K^\prime \frac{\varpi_1\varpi_2}{r_{12}}
\end{equation}
respectively,
where
\begin{equation}
    p_\imath:=\left|\bm{p}_\imath\right|,\;
r_\imath:=\left|\bm{r}_\imath\right|,\;
\bm{r}_{12}:=\bm{r}_1-\bm{r}_2,\;
r_{12}:=\left|\bm{r}_{12}\right|
\end{equation}
and $K'=K$ for the repulsive case,
$K'=-K$ for the attractive case,
and~$K'=0$ for the case of noninteracting orbiting bodies.
In all these cases both bodies are attractive to the center regardless of how they interact with each other. Dynamics for this system 
is described by the (free) Lagrangian
$L_0=T-U$.
Here~$K$ are the proportionality constants:
$K$=G=6.674$\times$10$^{-11}$m$^3$kg$^{-1}$s$^{-2}$
for the attractive case of gravitation
and $K=8.987\times10^{9}$~N m$^2$ C$^{-2}$ for the repulsive case of two equally charged bodies in the electromagnetic case.

We solve the Lagrangian to obtain the equation of motion for both orbiting bodies. The acceleration vector for each of the two bodies is
\begin{equation}
\label{eq:eqofmotion}
 \ddot{\bm{r}}_\imath     :=\frac{\text{d}^2
 \bm{r}_\imath}{\text{d} t^2}
  =-K V\frac{\varpi_\imath}{m_\imath r_\imath^3}\bm{r}_\imath+K'\frac{\varpi_1\varpi_2}
    {m_\imath r_{12}^3}\bm{r}_{12},
\end{equation}
for the $\imath^\text{th}$ body.
We have provided the relevant background
for two orbiting bodies in a central potential. Specifically, we
have described a general mathematical approach for the two
orbiting bodies that represent both repulsive and attractive
systems. This unified description of the two orbiting bodies
can be used for studying the limiting case of neutral systems,
i.e., two mutually noninteracting orbiting bodies. Now we
have equations of motion in the absence of external forces.
\subsection{External forces}
\label{external-forces}
In this subsection, we discuss external forces applied to the
two-body system in a central potential. We provide theoretical
and experimental background when the external forces are
applied to the repulsive or attractive and noninteracting two
bodies orbiting in a central potential. We present a mathematical description of the external force applied to one or both
orbiting bodies.

Examples of the external forces applied to two orbiting
bodies that are mutually repulsive have been explored for
argon and helium atoms, specifically in the case of ionization.
In this case, the external forces are laser pulses. Theoretically,
linearly~\cite{Yue}, circularly~\cite{Justin} and elliptically~\cite{Yueming} polarized laser pulses are used to study ionization processes in argon:
classical equations of motion are employed with averaging over a large ensemble of pairs of electrons in these cases.
Elliptically polarized~\cite{Wei_Wei} laser pulses are used in a helium atom to study the ionization process.
Experimentally,
elliptically polarized pulses are applied to study the strong field ionization of argon~\cite{pfeiffer}.

We provide examples of the case of two mutually attractive
bodies in a central attractive potential subject to an external
force. In the planetary system, electromagnetic thrust force
has been widely discussed for decades~\cite{Tsander}, used to make interstellar flight possible and to propel the spacecraft with $10\%$ of light speed~\cite{Young}.
However, experimental progress either in the laboratory or in the space environment is slow compared with theoretical progress~\cite{Forward}.
Recently, a photon thruster capable of amplifying the thrust was successfully used during the laboratory settings~\cite{Bae1,Bae2}, and a successful space implementation of solar sail was achieved.

We introduce the mathematical description of an external force and the shape of its envelope function applied
to repulsive, attractive, and noninteracting orbiting bodies.
Mathematically, such an external force is ~\cite{Wei_Wei}
\begin{align}
\label{eq:efield}
  \bm{F}\left(t\right)
        :=F_0f\left(t\right)\left(\cos\left(\omega_0 t\right)\hat{\bm x}+\epsilon\sin\left(\omega_0 t\right)\hat{\bm y}\right)
\end{align}
in general form,
for~$F_0$ the field amplitude coefficient, 
$\omega_0$ the angular frequency of this field,
$\epsilon$ the field ellipticity
and~$f(t)$ the temporal envelope function. The envelope function $f\left(t\right)$ changes slowly on the $T_0:=2\pi/\omega_0$
time scale
but is quite free in form otherwise.
We have ignored spatial dependence in Eq.~(\ref{eq:efield});
of course any applied force has spatial variation, but we focus on two extreme cases,
namely identical force applied to both orbiting bodies and the force applied only to one body and not to the other.

\begin{figure}
	\includegraphics[width=1\linewidth]{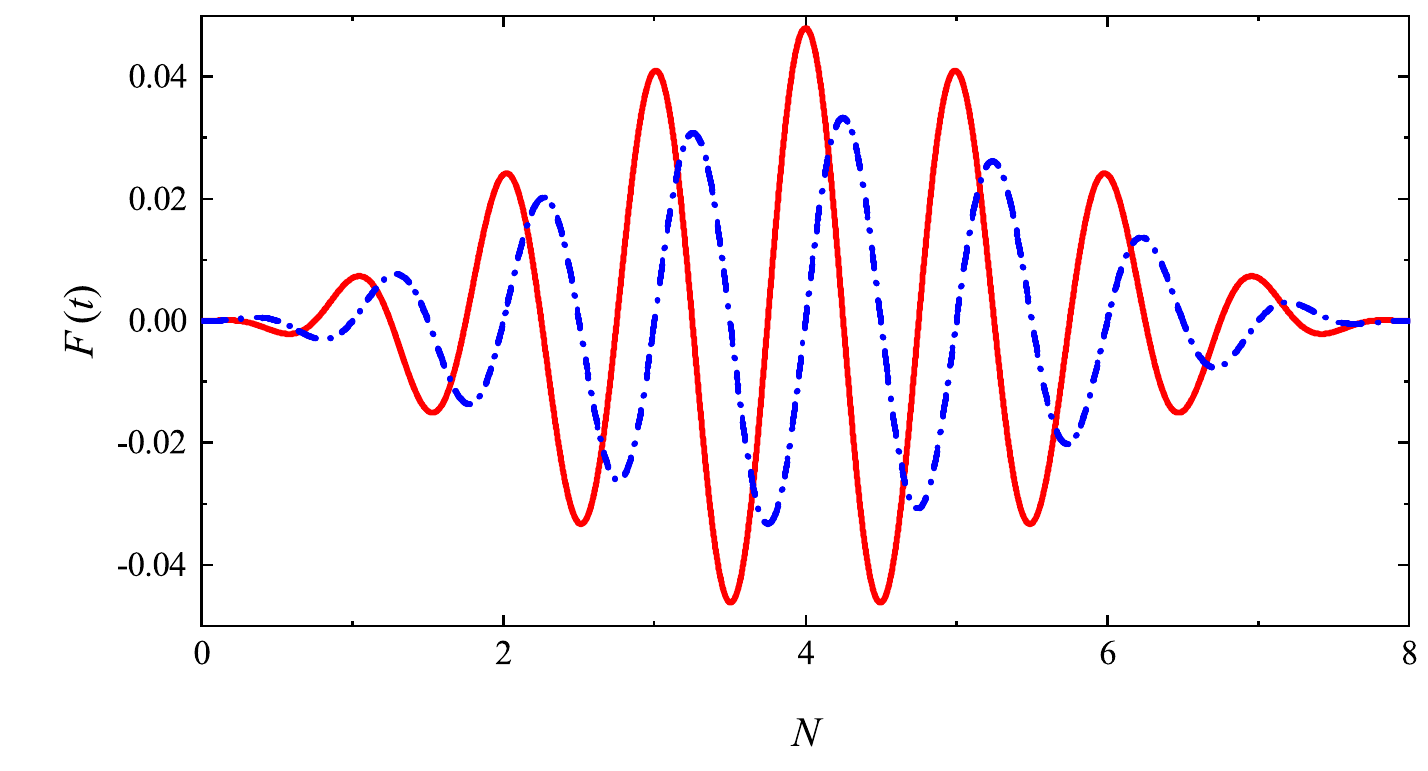}
\caption{Sine-squared shape of the external force [$\bm{F} (t)$ in a.u.] is plotted vs number of cycles~$N$
for the $\hat{\bm x}$
(red solid line)
and $\hat{\bm y}$
(blue dash-dot line)
components of the field,
respectively.}
	\label{fig:1}
\end{figure}

We provide some examples of the pulse shape that is widely used in repulsive orbiting bodies. Commonly used pulse shapes include trapezoidal used in nonsequential double ionization of He~\cite{Huang},
Gaussian in sequential double ionization of Ar~\cite{Yueming} and sine squared in nonsequential double ionization of Ar~\cite{Yue}, represented as 
\begin{equation}
\label{eq:pulseshape}
   f\left(t\right)=\sin^2\left(\frac{\omega_0 t}{2N}\right)
\end{equation} 
for~$N$ the number of cycles. Figure~\ref{fig:1}
shows this sine-squared shape external force. In the next section, we describe the effects of this external force on the orbiting bodies.
\subsection{Creating collisions}
\label{subsec:creatingcollisions}
In this subsection, we discuss two-body collisions in mutually repulsive, mutually attractive, and noninteracting orbiting bodies in a central potential. Theoretical and experimental
background are introduced for the case that external forces
create a collision between mutually repulsive or mutually
attractive orbiting bodies.

Now we explain the two-body collisions orbiting in a central potential. Collisions are significant scattering events that correspond to a large change of trajectory in a short time.
Deflection is the change of the body’s velocity as a result
of collision. Rutherford scattering~\cite{Rutherford} is a special case of repulsive forces between bodies. 

We provide examples of the two-body collisions in mutually repulsive and mutually attractive bodies orbiting in a central potential.
In a mutually repulsive two-body system, the electron-electron collision is responsible for He$^{+2}$ ion studied experimentally~\cite{Fittinghoff,Walker} and theoretically~\cite{Yue,Chen} in the non-sequential double ionization process of the two-body atomic system.
The geometric parametrization method has been introduced to study the two elastic collisions of planetary bodies~\cite{Rica}. 
In planetary system, the collision of two planetary bodies is used by various spacecraft to reach their target~\cite{papkov} and to calculate the trajectories of spacecraft between two planets~\cite{RicadaSil}.

One excellent approach to launching orbiting bodies is to exploit the slingshot, or gravity assist,
method~\cite{Strange}.
The gravity-assist maneuver exploits relative motion between the body being launched and another body that increases or decreases the speed of the body being launched and/or guides its direction.
Multiple gravity assist has proven to be valuable by saving propellant, time and expense~\cite{papkov}.
For example, the Galileo spacecraft
used gravity assist provided by Venus plus two assists from Earth to reach its destination of Jupiter~\cite{Fischer}.
Similarly, Cassini, Voyager 1 and Voyager 2 have used multiple gravity-assist maneuver
to help them reach their destination.

We summarize the background on existing methods to determine whether a collision has taken place. In one case, this
assessment is performed by comparing the velocity of the two
orbiting bodies before and after the collision. The event-drive
simulation approach is used in the case of spherical bodies
collision to connect the velocity of the bodies before and after
collision~\cite{luding}.
Another approach is to follow the trajectories of the colliding bodies by solving their equation of motion~\cite{Luding1998}. 

In this section, we discussed concepts, mathematics, and
methods that describe orbiting bodies, external forces, and
strong scattering events. In the next section, we proceed to describe our approach to solving forced deterministic collisions
between nonrelativistic orbiting bodies in a central potential.
\section{Approach}
\label{Sec:3}
In this section, we present our approach to explain the
two-body system for repulsive, attractive, and noninteracting
two orbiting bodies. We describe our model, its mathematical
representation, and our methods to solve numerically this two body system orbiting in a central potential, which is subject to
external forces. To validate our results, we confirm conservation of energy and momentum numerically.
\subsection{Model}
\label{subsec:model}
In this subsection, we introduce our model of the two-body
system orbiting in a central potential for three different cases,
i.e., repulsive, attractive, and noninteracting orbiting bodies.
We define a collision in such a way that the same term applies
in each case.

First we describe our model of two orbiting bodies in the
absence of an external force. By assuming spherical orbiting
bodies, we neglect multipolar terms. As we treat bound orbiting bodies, at least prior to the controlled collision, only elliptical orbits are treated as initial conditions. For simplicity,
we treat the case that the two orbiting bodies are corotating,
coplanar, and concentric orbits with different eccentricities.
The two bodies orbit a central potential nonrelativistically,
and we specifically focus on the simple case of the inverse
square-law force, which is readily generalizable. We prefer
the Lagrangian approach because constraints are easier to
accommodate than for the Hamiltonian.

We consider two bodies orbiting in a central potential in the
presence of an external force. In general, forces acting on the
two orbiting bodies can be independent and moreover vary in
time and in space. We consider two extreme cases pertinent
to the examples we consider: creating controlled collisions
between two highly excited electrons in a heliumlike atom
and creating controlled collisions between planetary bodies
orbiting in a gravitational field. In the former case, the force
would be an electromagnetic field with a carrier wavelength,
constrained by near-resonance conditions with orbiting frequencies, that far exceeds the size of the atom. Thus the
driving force is treated as identical for both electrons and
constant in space. For the gravitational case, we consider a
thrust mechanism attached to the orbiting body, in which case
the force is time dependent but spatially independent and acts
only on the inner body and not the outer body. In the electronic
case, magnetic interactions are regarded as negligible in the
case of the electromagnetic field.

Now we discuss how to choose the shape of the external
force function. Based on numerical analysis of alternative
external force function for an elliptical orbit, the sine-squared
force emerged as the best~\cite{Heslar}. We thus adopt this sine-squared force function as well following this positive result. 

Here we explicitly define what we mean by collision. The
concept of a collision is introduced when the lower orbiting
body, subject to external force collides with the higher orbiting
body. We define a collision in such a way that the same
term applies to both the atomic (electron-electron collision)
and planetary system (Earth-Mars collision). To define the
collision, the symbols $E$ and $\tau$ represent the energy and time period of the lower orbiting body, respectively.

Now we discuss a model of two orbiting bodies that are
mutually repulsive. Specifically, in the repulsive case, we
treat the helium atom as the representative two-body atomic
system. The classical decay of the orbits is neglected because
the duration of the laser field is faster than the time of decay.
Although Rydberg states were introduced for hydrogen, the
Rydberg states for high principal quantum number orbits in
two-electron atoms are analyzed~\cite{Agner}.
In particular, we consider two sets of high Rydberg states having principal quantum numbers ($n_\text{i}=20$, $n_\text{f}=30$) and ($n_\text{i}=100$, $n_\text{f}=110$).
In these cases the orbiting bodies are attracted to the central body and repel each other.

Now we explain how the precision of our classical description is expected to scale with principle quantum number~$n$
by using underlying principles of the Bohr atom;
the scaling rules we obtain here provide a guide for how accurate the classical model is,
but of course a full quantum mechanical treatment is needed to achieve highly accurate predictions and well designed control sequences.
Consider an electron in a highly excited orbital represented by~$n$.
In atomic units,
we consider that the Bohr radius is~$n^2$,
and the momentum (which is the same as velocity in atomic unites) is~$1/n$.
We describe the dependences of both position and momentum variances on principle quantum number $n$.
If we fix position uncertainty to be $n^{3/2}$ and fix the momentum uncertainty to be $n^{-3/2}$,
the localized electron's position and momentum satisfy the Heisenberg uncertainty product relation so is in a minimum-uncertainty state.
Furthermore,
the fractional uncertainty of position is $n^{3/2}/n^2=1/\sqrt{n}$,
and, similarly,
the fractional uncertainty of momentum is $n^{-3/2}/(1/n)=1/\sqrt{n}$.
Thus the position and momentum uncertainties are the same.
For $n=100$, the uncertainty, hence error, is about 10\%, and the classical description thus has an accuracy that improves with the square root of the principle quantum number.
A semiclassical description, perhaps using Perelomov coherent states,
could possibly provide a perturbation-based correction method to offset this error.

We consider a model of mutually attractive and noninteracting two orbiting bodies. In the attractive interacting and
noninteracting cases, we consider planetary systems (Earth
and Mars) orbiting the Sun. In the attractive interacting case,
the orbiting bodies have attractive interaction with a central
body and with each other, whereas in the noninteracting case
we only consider attractive interaction of the orbiting bodies
with the central body and neglect any other interaction between the orbiting bodies.

We validate our model by checking that energy and
momentum are conserved before and after the collision.
Specifically, we consider two orbiting bodies, each with energy and momentum at the initial time $t=0$,
and we confirm
that, within a reasonable error tolerance, the resultant total
energy and momentum for the two bodies, which might or
might not still be in orbit after the collision, satisfy the conservation law at the final time~$t^\prime$.
As the collision is brought
about by an external force, and this external force itself has
energy and momentum, we include this external energy and
external momentum in the calculation.
\subsection{Mathematics}
\label{sec:mathematics}
In this subsection, we present the mathematics for a two body system orbiting in a central potential for an external
force applied on one or both bodies. To validate our results,
we check that energy and momentum are approximately conserved before and after the collision.

We present a mathematical description of the two-body
system for an external force applied to one (lower) or both
orbiting bodies. The total two-orbiting-body Lagrangian is
\begin{equation}
\label{eq:Lagrangian}
 L=L_0+\sum_{\imath=1}^2\bm{r}_\imath\cdot\bm{F}_\imath\left(t\right),
\end{equation}
with~$L_0$ the time-independent free (or ``drift," in the language of control theory~\cite{Arenz}) term
and~$\bm{F}_\imath\left(t\right)$
the time-dependent control force~(\ref{eq:efield}) acting on the $\imath^{\text{th}}$ body.
For the force acting only on the lower body,
we impose $\bm{F}_2(t)\equiv\bm0$,
and, for the case that identical forces act on both bodies,
we impose $\bm{F}_1(t)\equiv\bm{F}_2(t)\equiv\bm{F}(t)$.
Motion occurs in a two-dimensional plane with coordinates~$x$ and~$y$,
and we employ unit Cartesian coordinates~$\hat{\bm{x}}$ and
$\hat{\bm{y}}$.

Now we describe how the Lagrangian Eq.~(\ref{eq:Lagrangian}) is solved by using the Euler-Lagrange equation of motion when the external force is either coupled to the lower orbiting body or else coupled to both bodies for the case of identical forces applied to both.
In the case that the force is applied only to the lower body,
we obtain coupled second-order differential equations of motion, as
\begin{equation}
\label{eq:acc1}
 \ddot{\bm{r}}_1
    =-K\frac{V \varpi_1}{m_{\varpi_1} r_1^3}\bm{r}_1+K^\prime\frac{\varpi_1 \varpi_2}
    {m_{\varpi_1} r_{12}^3}\bm{r}_{12}-\frac{1}{m_{\varpi_1}}\bm{F}(t),
\end{equation}
and
\begin{equation}
\label{eq:acc2}
 \ddot{\bm{r}}_2
    =
  -K\frac{V \varpi_2}{m_{\varpi_2} r_2^3}\bm{r}_2+ K^\prime\frac{\varpi_1 \varpi_2}
    {m_{\varpi_2} r_{12}^3}\bm{r}_{12}.
\end{equation}

Similarly, we can solve the Lagrangian~(\ref{eq:Lagrangian}) for the case that an external force is coupled to both orbiting bodies. In this case, we obtain the second-order differential equations of motion, as
\begin{equation}
\label{eq:acc3}
\ddot{\bm{r}}_1
    =-K\frac{V \varpi_1}{m_{\varpi_1} r_1^3}\bm{r}_1+ K^\prime\frac{\varpi_1 \varpi_2}
    {m_{\varpi_1} r_{12}^3}\bm{r}_{12}-\frac{1}{m_{\varpi_1}}\bm{F}(t),
\end{equation}
and
\begin{equation}
\label{eq:acc4}
\ddot{\bm{r}}_2
    =-K\frac{V \varpi_2}{m_{\varpi_2} r_2^3}\bm{r}_2+ K^\prime\frac{\varpi_1 \varpi_2}
    {m_{\varpi_2} r_{12}^3}\bm{r}_{12}-\frac{1}{m_{\varpi_2}}\bm{F}(t).
\end{equation}
These equations of motion (\ref{eq:acc1})--(\ref{eq:acc4}) apply to both cases of repulsive and attractive forces,
which extends standard descriptions of classical motion by incorporating force terms manifested as laser driving fields in the atomic case and as thrust terms in the gravitational case.

These equations of motion for the two orbiting bodies accounts for collisions when the two bodies coincide.
We formalize the definition of collision as follows,
accounting for near
(quantified by small parameter~$\varepsilon$)
impact rather than direct impact in the case of repulsive interaction.
\begin{definition}
For some~$\varepsilon\in(0,1)$,
a collision of a body with energy~$E$
over duration~$\tau$
is a fractional energy change $\frac{\delta E}{E}$ during fractional time $\frac{\delta \tau}{\tau}$ such that $\frac{\delta E}{E}<\varepsilon$ and $\frac{\delta \tau}{\tau}>\varepsilon$.
\end{definition}
\begin{remark}
For an orbiting body, the fractional time is with respect to period~$\tau$.
\end{remark}

The fractional energy change of an orbiting body $\frac{\delta E}{E}$ for $E$ is the instantaneous energy of the body at time~$t$.
The fractional time scale of the collision is $\frac{\delta \tau}{\tau}$ for $\tau$ the period of the orbit.

Now we describe how we validate our calculations by checking that conservation of energy and momentum before and after the collision holds.
At the initial time ($t=0$)
and the final time ($t=t^\prime$),
the total two-body energy is $\sum_{\jmath=1}^2 E_{\jmath}(t)$
for~$E_{\jmath}$ the energy of the $\jmath^\text{th}$ body.
For the $\jmath^\text{th}$ body,
its position~$\bm{r}_\jmath$
and velocity~$\dot{\bm{r}}_\jmath$
are specified;
the corresponding total energy is
\begin{equation}
\label{eq:jthenergy}
   E_\jmath
        =\frac{p_\jmath^{2}}{2 m_\jmath}-K\frac{V \varpi_\jmath}{r_\jmath} +K' \frac{\varpi_1\varpi_2}{r_{12}}
\end{equation}
at any time~$t$.
The total energy gain by application of the external force~(\ref{eq:efield}) is
\begin{equation}
\label{eq:extenergy}
E_\text{ext}
    =\int_0^{t^\prime}\text{d}t\,\dot{\bm{r}}\cdot\bm{F}(t).
\end{equation}
Mathematically, the fractional error in energy~fe$(E)$ for the pair of orbiting bodies is
\begin{equation}
    \label{eq:feE}
  \text{fe}(E)
    =\left|\frac{E_\text{final}-E_\text{initial}}{E_\text{final}+E_\text{initial}}\right|,  
\end{equation}
where
\begin{equation}
\label{eq:energychange}
   E_\text{final}=\sum_{\jmath=1}^2 E_{\jmath}(t^\prime), \quad E_\text{initial}=\sum_{\jmath=1}^2 E_{\jmath}(t=0)+E_{\text{ext}}.
\end{equation}
This quantity shows the error because energy is conserved so a nonzero value constitutes an error. Here we are comparing before vs after the collision but this relation is valid for any comparison of before and after any event.

Similarly,
we test conservation of vector momentum
for collision between the two bodies with
\begin{equation}
\label{eq:jthmomentum}
   \bm{p}_\jmath
   =m_{\varpi_\jmath}\dot{\bm{r}}_\jmath,
\end{equation}
the momentum vector of the $\jmath^\text{th}$ body.
The total momentum gain,
due to application of the external force,
is
\begin{equation}
\label{eq:extmomentum}
    \bm{p}_\text{ext}
        =\int_0^{t^\prime}\text{dt}\bm{F}(t).
\end{equation}
The fractional error for momentum is \begin{equation}
\label{eq:fep}
    \text{fe}(p)
    =\left|\frac{p_\text{final}-p_\text{initial}}{p_\text{final}+p_\text{initial}}\right|,
\end{equation}
where
\begin{equation}
\label{eq:momentumchange}
\bm{p}_\text{final}=\sum_{\jmath=1}^2 \bm{p}_{\jmath}(t^\prime), \quad \bm{p}_\text{initial}=\sum_{\jmath=1}^2 \bm{p}_{\jmath}(t=0)+\bm{p}_{\text{ext}},
\end{equation}
with this quantity representing the fractional error because momentum should be conserved.
\subsection{Methods}
\label{subsec:methods}
Now we describe our method to solve acceleration~(\ref{eq:acc1}) and~(\ref{eq:acc2}) for three different cases, i.e., repulsive, attractive and noninteracting two orbiting bodies.
We solve acceleration~(\ref{eq:acc3}) and~(\ref{eq:acc4}) for an external force acting on both orbiting bodies.
We explain our numerical method for solving the problem
for each case, and we explain how we validate our results
by applying other numerical methods to check consistency.
Furthermore, we check that energy and momentum are approximately conserved for before and after the collision.

Now we explain our primary numerical technique for solving the acceleration equations, and then we explain the other
numerical techniques that we use to validate by confirming
that the same answers are obtained. As we are solving a
pair of coupled second-order differential equations over two-dimensional vector quantities, we opt to use the well tested
and reliable Runge-Kutta (4,5) method~\cite{John}.
For this technique, we first convert our second-order differential equations
into equivalent first-order equations. Then we use the initial
values, namely, positions and velocities of the two bodies,
as inputs to solve these equations for our two cases of a
two-electron atomic system~\cite{Panfili} and for a planetary system~\cite{Planetary}.  

The numerical Runge-Kutta (4,5) method  simultaneously solves the equations of motion with local orders of both~4 and~5 using two Runge-Kutta procedures~\cite{Reichelt}.
We opt for the Runge-Kutta (4,5) method as this method works for the following criteria:
the differential equations have smooth solutions and
high accuracy is needed.
The Runge-Kutta (4,5) method is built into MATLAB\textsuperscript{\textregistered},
and we use temporal discretization to obtain a feasible solution.

We use three other numerical methods to validate our results.
Our first alternative is the explicit Runge-Kutta (2,3) method~\cite{bogack}.
Our second validation technique uses the variable-step, variable-order Adams-Bashforth-Moulton PECE solver~\cite{Lawrence}.
Our final validation technique employs Gear's Method~\cite{shampine}. We use the same discretization time in all four cases to compare the results for these numerical methods.

Now we explain our method for calculating and validating
approximate conservation of energy and momentum for the
two-body colliding system. Given initial position and velocity
we solve Eqs.~(\ref{eq:jthenergy}) and~(\ref{eq:jthmomentum}) to calculate the initial energy and momentum of the two orbiting bodies.
Then we calculate the external force function~(\ref{eq:efield}) and thence the external
energy~(\ref{eq:extenergy}) and momentum~(\ref{eq:extmomentum}) applied to the inner orbiting body.
Subsequently,
we compute the final energy and momentum by solving the positions and velocities
of the two orbiting bodies
at the final time~$t^\prime$.

Based on initial and final momentum and energy, we check whether the conservation laws have been fully respected by our numerical calculations. 
The fractional error in energy~(\ref{eq:feE}) and in momentum~(\ref{eq:fep}) is calculated by each numerical method. The validation holds if these four numerical results provide sufficient convergence. 

In conclusion, we have introduced our model, mathematics, and method for solving a two-body system orbiting in a
central potential by considering three cases, namely, repulsive,
attractive, and noninteracting orbiting bodies. Our Methods
subsection makes clear that the equations are solved in the
same way regardless of which of these three cases applies.
In the next section, we present the results for the repulsive,
attractive, and noninteracting two orbiting bodies which are
subject to external forces.
\section{Results}
\label{Sec:4}
In this section, we present results for our three cases,
namely, repulsive, attractive, and noninteracting two orbiting bodies. We present these results for specific examples,
namely, the helium atom for the repulsive case and Earth and
Mars for attractive and noninteracting cases. Causing Earth
to collide with Mars is an unlikely, and also undesirable, scenario, but this example allows us to illustrate this controlled
collision in the context of broadly understood parameters. We
validate our results by calculating conservation of energy and
momentum. 
\subsection{Repulsive interaction: Collision between bound electrons in an atom}
In this subsection, we present the results for two repulsively interacting orbiting bodies (electrons). The two
orbiting electrons interact attractively with the central body
(nucleus) and interact repulsively with one another. Controlled
collisions between electrons are created by applying a time-dependent external force (electromagnetic field) and we depict the results for two cases: with and without the external force acting on the second electron.

\begin{figure*}
\includegraphics[width=0.8\linewidth]{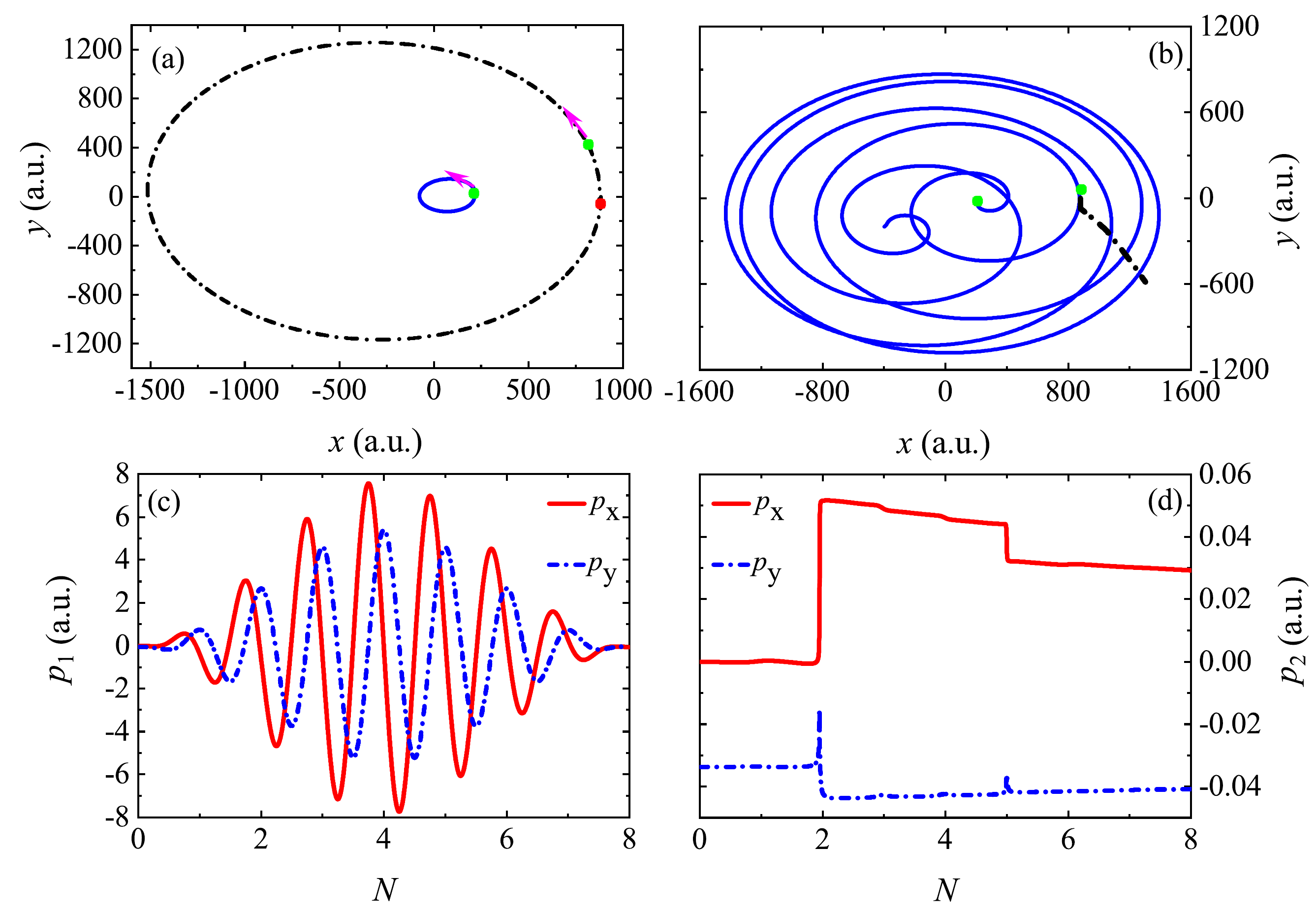}
\caption{%
\textbf{Repulsive interacting case
with an external force applied only to the inner body:}
(a)~counterclockwise
elliptical orbits of two bodies
(solid blue line for the inner orbit and black dash-dot line for the outer orbit)
with
green dots signifying initial positions,
the red dot signifying the collision point, and
magenta arrows signifying directions of the orbits
and the positions in the orbital plane,
all quantified in atomic units.
(b)~Trajectories for the inner orbiting body (blue solid line)
and outer orbiting body (black dash-dot)
commencing at the positions signified by the green dots at the moment when the external field is turned on,
which continues for eight cycles,
and shows a collision after two cycles,
and the units of position are in atomic units.
Two components of the two-dimensional momentum vector
(red solid line for the $x$ component and blue dash-dot line
for the $y$ component),
in atomic units,
as a function of time counted in terms of the number of cycles~$N$
for (c)~the inner orbiting body
and (d)~the outer orbiting body.
Values for angular frequency and field amplitude are $\omega_0=0.005$ (a.u.) and $\text{F}_0=0.044$ (a.u.),
respectively.}
	\label{fig:repres}
\end{figure*}

In the absence of any external field, we describe the initial
position of the two orbiting electrons as shown in Fig.~\ref{fig:repres}(a). Initially, electrons are orbiting in two different orbits as defined in~Sec.~\ref{subsec:model}. The time period for moving electrons in the Rydberg states is well described by Wang and Robicheaux~\cite{Xiao}.

Our result for the repulsive two-body system,
with an external force~(\ref{eq:efield}), applied only to the inner orbiting body (at $n_\text{i}=20$),
causing the inner orbiting body to rise to the outer orbiting body's orbit (at $n_\text{f}=30$) for a collision,
is presented in Fig.~\ref{fig:repres}(b).
For this case,
the applied laser field has $N=8$ cycles 
with field amplitude, field ellipticity and angular frequency set at $\omega_0=0.005$ (a.u.), $\epsilon=0.7$ and $\textit{F}_0=0.044$ (a.u.),
respectively.
We see clearly in Fig.~\ref{fig:repres}(b)
that the laser field launches the inner orbiting body into a spiral that reaches the outer orbiting body's
higher orbit and collides with this outer orbiting body.

Figure~\ref{fig:repres}(b) shows clearly the complex dynamics leading
to the collision. The outer electron moves in an elliptical orbit
prior to the collision, but this elliptical orbit is not evident due
to the driving force being so fast on the outer orbital time scale
that the outer orbiting body has barely moved in the depicted
time scale. We see the collision leading to the outer orbiting
body suddenly adopting a new trajectory that takes the body
to an unbound state, and the inner orbiting body changes its
motion, falling back to a lower orbit instead of going to a
higher orbit.

We can see a huge change of momentum for the outer
orbiting body upon collision and a small change of momentum
for the inner orbiting body due to the collision. These time-dependent momenta are depicted in Figs.~\ref{fig:repres}(c) and~\ref{fig:repres}(d)
for the inner and outer bodies, respectively.
The $x$ and $y$ momenta are shown for each body as a function of the number of cycles~$N$.
The external driving force~(\ref{eq:efield}) is stronger along the $\hat{\bm x}$ axis than along the $\hat{\bm y}$ axis,
and this effect is clearly shown by the fact that the~$p_\text{x}$ locus has larger amplitude than the~$p_\text{y}$ locus for both bodies,
as depicted in Figs.~\ref{fig:repres}(c) and~\ref{fig:repres}(d).
Furthermore, the loci of the inner orbiting body shown in Fig.~\ref{fig:repres}(c) matches closely the locus of the driving force shown in Fig.~\ref{fig:1}.
Two kinks in the momenta of the outer orbiting body are evident in Fig.~\ref{fig:repres}(d) at the second and fifth cycle
due to a collision and a recollision,
respectively,
between the inner and the outer bodies.
These collisions are implicit in Fig.~\ref{fig:repres}(b);
i.e., the trajectories of the two bodies involve two crossing points that represent collisions,
and the other crossing points are ``misses".
\begin{figure*}
\includegraphics[width=0.8\linewidth]{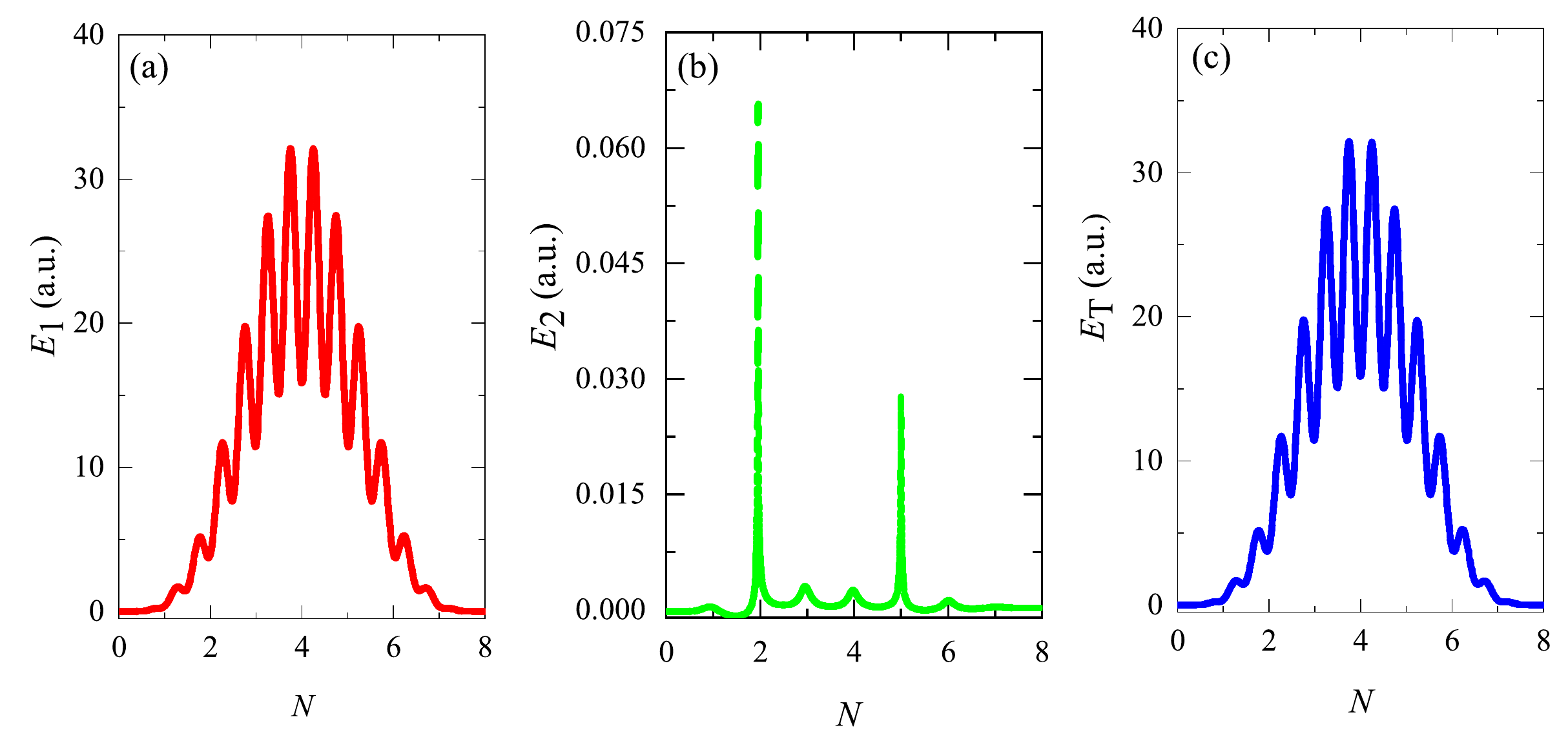}
\caption{\textbf{Repulsive interacting case
with an external force applied only to the inner body:}
electron energy vs number of cycles~$N$
for (a)~energy of the inner orbiting body,
(b)~energy of the outer orbiting body and 
(c)~total energy (both bodies plus the driving field).%
}
	\label{fig:repenergy}
\end{figure*}

Here we calculate the energy for the repulsive interacting case with an external force applied only to the inner body,
discussed in~Sec.~\ref{sec:two-orbiting-bodies},
of each of the two orbiting bodies and present the energy of each of the bodies and the sum energy in Figs.~\ref{fig:repenergy}(a),~\ref{fig:repenergy}(b) 
and~\ref{fig:repenergy}(c),
respectively,
over $N=8$ cycles.
Although not evident in the plot,
the energy of each body at $N=0$
is negative,
due to the bodies being bound initially;
at $N=8$,
again not evident in the figure due to scale,
the initial body's energy is negative,
signifying being bound,
whereas the outer orbiting body's energy is positive,
signifying that is has escaped the binding potential.
In Fig.~\ref{fig:repenergy}(b),
we can see more than two effective collisions:
the two major peaks are the collision and recollision discussed for Fig.~\ref{fig:repres}(d),
and the other peaks are smaller collisions arising due to longer-range repulsive interactions.
\begin{figure*}
\centering
\includegraphics[width=0.8\linewidth]{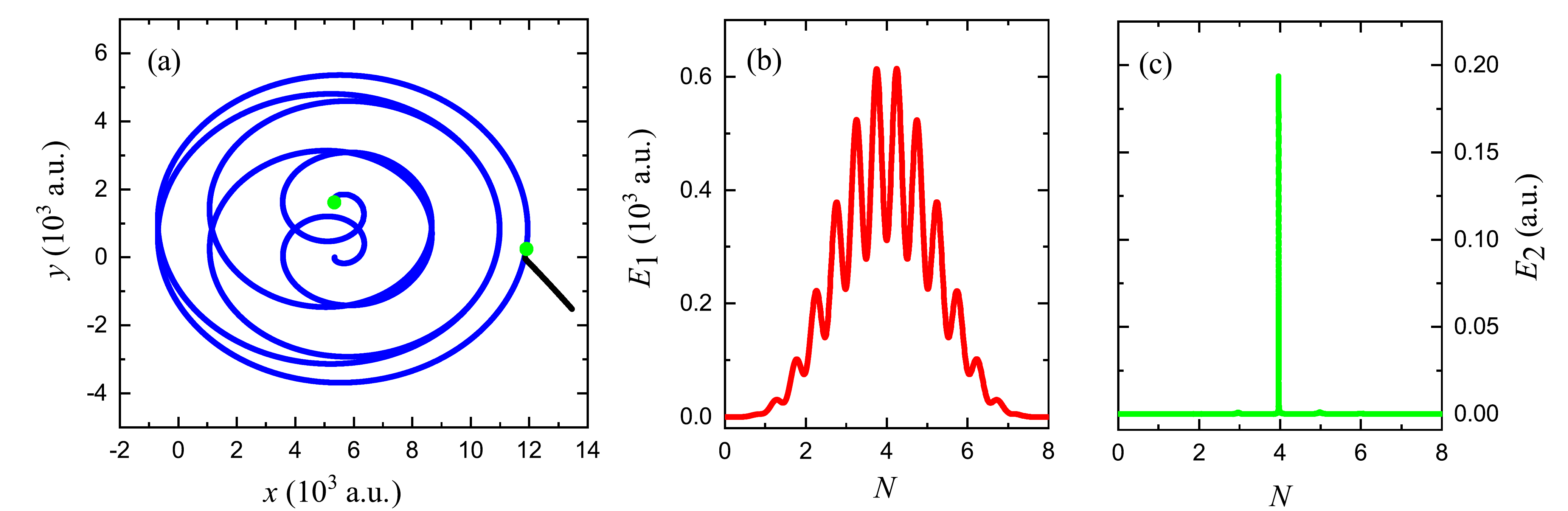}
\caption{\textbf{Repulsive interacting case
with an external force applied only to the inner body:}
(a) trajectories for the inner orbiting (blue solid line) and outer orbiting body (black solid line) commencing at the positions signified by the green dots at the moment when the external field is turned on, which continues for eight cycles, and shows a collision after four cycles. Electron energy vs number of cycles $N$ for (b) energy of the inner orbiting body and (c) energy of the outer orbiting body.%
}
	\label{fig:repn100}
\end{figure*}

Now we present the results for the repulsive two-body system when the inner and outer orbiting bodies are initially at $n_{\text{i}}=100$ and $n_{\text{f}}=110$, respectively. The external force~(\ref{eq:efield}) is applied to the inner orbiting body, causing the inner orbiting body to rise to the outer orbiting body's orbit for a collision, as shown in Fig.~\ref{fig:repn100}. In this case, the field amplitude value is $\textit{F}_0=0.1948$ (a.u.), whereas all other parameters are the same as in Fig.~\ref{fig:repres}.

Figure~\ref{fig:repn100}(a) shows clearly that the external field launches the inner orbiting body into a spiral that reaches the outer orbiting body's higher orbit and collide with this outer body. Prior to collision, the outer orbiting body moves in elliptical orbit which is not evident in Fig.~\ref{fig:repn100}(a), because the driving force is so fast on the outer orbital time scale that the outer orbiting body has barely moved in the depicted time scale. The outer orbiting body has adopted a new trajectory as a result of the collision, which takes the body to an unbound state, and the inner orbiting body changes its movement, falling back to a lower orbit.

Figures~\ref{fig:repn100}(b) and~\ref{fig:repn100}~(c) show the energy results for repulsive orbiting bodies with an external force applied only to the inner orbiting body over $N=8$ cycles. Although not evident in the plot, the energy of each orbiting body at $N=0 $ is negative, meaning that the orbiting bodies are bound. At $N=8$, again not evident in the figure due to scale, the inner body's energy is negative, signifying being bound, whereas the outer orbiting body's energy is positive, signifying that it has escaped the binding potential. In Fig.~\ref{fig:repn100}(c), the peak shows the energy gain of the outer orbiting body during the collision.
\begin{figure*}
\includegraphics[width=0.8\linewidth]{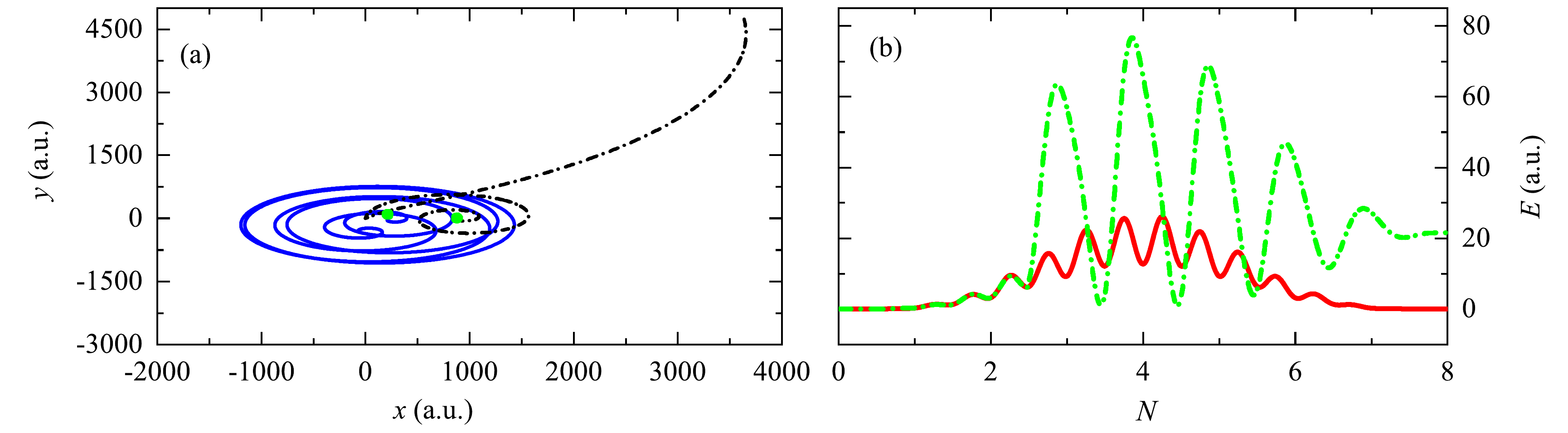}
\caption{\textbf{Repulsive interacting case with an external force applied to both bodies:}
(a)~trajectories for the inner orbiting body (blue solid line)
and outer orbiting body (black dash-dot)
commencing at the positions signified by the green dots at the moment when the external field is turned on,
which continues for eight cycles,
and shows a collision after two cycles,
and with the units of position in atomic units.~(b)~Electron energy vs number of cycles~$N$ 
for the inner orbiting body (red solid line) and for the outer orbiting body (green dash-dot line).}
	\label{fig:repulsive2}
\end{figure*}

We now present the results for the repulsive two-body system, with an external force~(\ref{eq:efield}) applied to both orbiting bodies to create the collision as shown in Fig.~\ref{fig:repulsive2}(a). For this case,
the applied laser field has $N=8$ cycles 
with field amplitude and field ellipticity $\textit{F}_0=0.04$ (a.u.), whereas all other parameters are the same as in Fig.~\ref{fig:repres}. Figure~\ref{fig:repulsive2}(a) clearly shows that the laser field launches both two orbiting bodies into a spiral motion away from the center.
The two bodies collides with each other after the second cycle.
After collision, the outer orbiting body suddenly adopts a new trajectory that takes the body to an unbound state,
i.e., escapes the system,
and the inner orbiting body still moves like a spiral.

Figure~\ref{fig:repulsive2}(b) depicts the energy for the repulsive interacting case with an external force applied to both bodies, described in Sec.~\ref{sec:two-orbiting-bodies}, of each of the two orbiting bodies over $N=8$ cycles. Although not evident in the plot, the energy of each body at $N=0$ is negative, due to the bodies being bound initially; at $N=8$, again not evident in the figure due to scale, the initial body's energy is negative,
signifying being bound, whereas the outer orbiting body's energy is positive, signifying that it has escaped the binding potential.

In this subsection we have explained a specific application of our repulsive-interaction framework by analyzing a classical model for highly excited electrons in a helium atom.
We describe the initial conditions and the external force applied to one or both electrons and show their dynamics until and beyond collision.
Then we study momenta and energy and describe features of the resultant plots.
The next two subsections follow the same structure.
\subsection{Attractive interaction: Collision between orbiting bodies in a gravitational central potential}
In this subsection, we present our results
for two attractively interacting orbiting bodies
in a central potential perhaps created by a highly massive body such as the Sun.
As a concrete example,
we consider our planetary system
with Earth as the inner orbiting body and Mars as the outer body orbiting the Sun,
as an application of our technique.
\begin{figure*}
 	\includegraphics[width=0.8\linewidth]{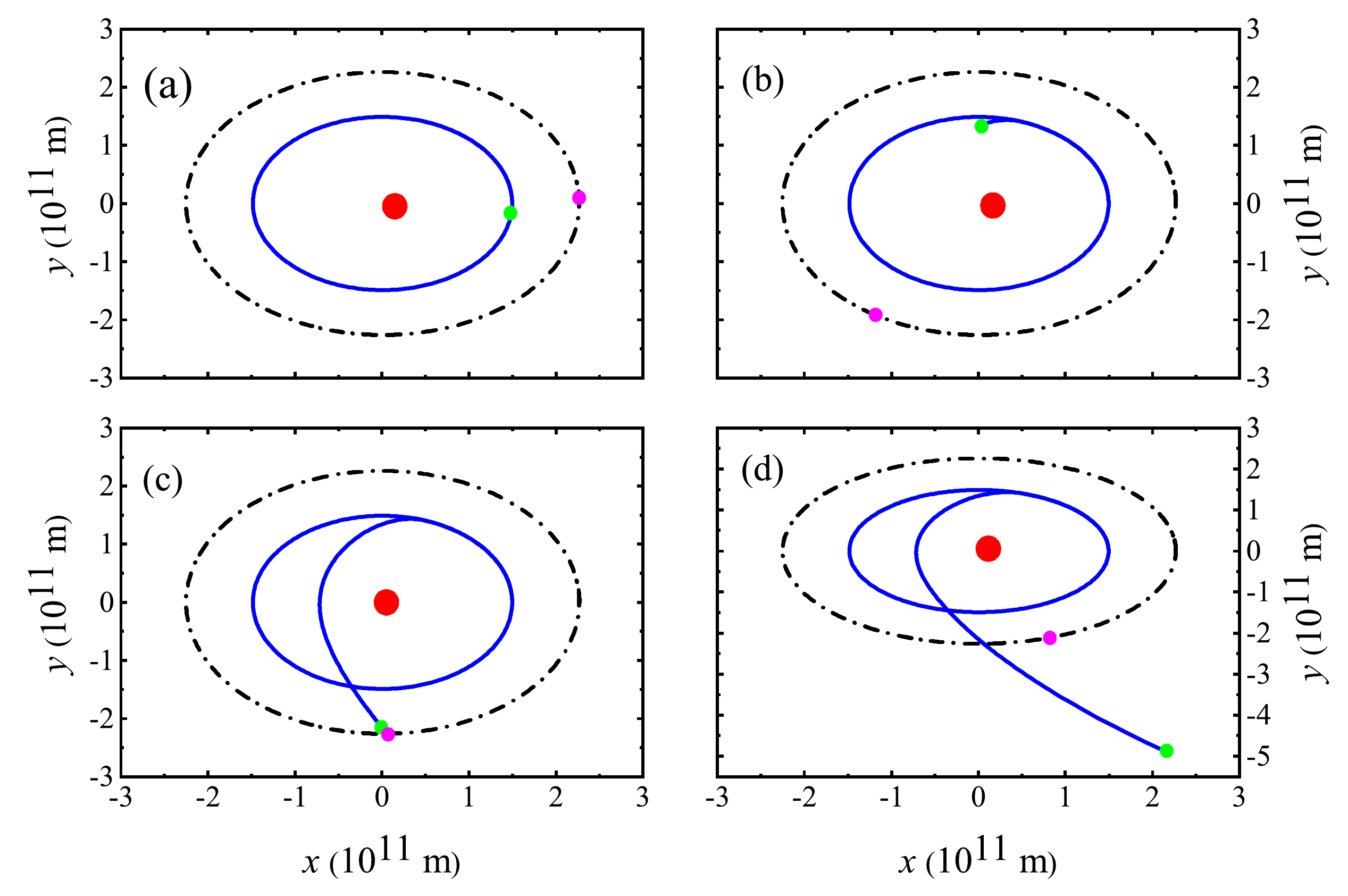}
\caption{\textbf{Attractive interacting case:}
elliptical orbits of Earth (the green dot signifies Earth in a given time and the solid blue line is the trajectory of Earth's orbit) 
and Mars (the magenta dot signifies Mars in a given time and the black dash-dot line is the trajectory of Mars's orbit); a red dot for the Sun at one focus for
(a)~Earth and Mars initial position at time $t=0$,
(b)~when the external thrust force acts on Earth,
(c)~gravity assist creates a collision between Earth and Mars and
(d)~Earth and Mars dynamics after the collision.}
	\label{fig:attracdynamics}
\end{figure*}
In the attractive interacting case, we explore our planetary case of Earth and Mars orbiting the Sun and ignore other bodies such as the moon and other planets.
Controlled collisions between planetary bodies are created by applying a time-dependent external force only to the lower orbiting body.

Concentric orbits with different eccentricities (0.017 for Earth and 0.093 for Mars) are considered.
As Earth completes one cycle,
the external thrust force is optimally applied at the time of 1.2 Earth years. 
The angular frequency~$\omega_0$
corresponds to the orbital frequency for Earth,
which equals one year to complete one cycle.
The external thrust force
$\text{F}_0\approx 10^{23}$~N,
with function shape shown in Fig.~\ref{fig:1},
is applied to the lower orbiting body.

Now we present dynamics for two attractively interacting orbiting bodies.
Figure~\ref{fig:attracdynamics}(a) shows the initial positions of the orbiting bodies at time $t=0$ when no external field is applied to the system.
Initially, the external field is off
but then is turned on after 1.2 Earth years.
We keep the thrust off for 1.2 years to show dynamics of Earth's orbiting motion and also wait for an opportune moment to launch Earth to a higher-energy state that takes Earth close to Mars.
At the launch time,
we simulate applying thrust to Earth when Mars is on the opposite side as shown in Fig.~\ref{fig:attracdynamics}(b).

In Fig.~\ref{fig:attracdynamics}(c)
we see the Earth orbit for 1.2 years and then move closer to the Sun before moving to a higher orbit.
This counterintuitive motion is a manifestation of gravity assist described in~Sec.~\ref{subsec:creatingcollisions}.
In this case the Earth is first pushed closer to the sun,
and the external thrust force plus the Sun's gravity assist launches Earth away from the Sun where it collides with Mars
as shown in Fig.~\ref{fig:attracdynamics}(c).

\begin{figure*}
 	\includegraphics[width=0.8\linewidth]{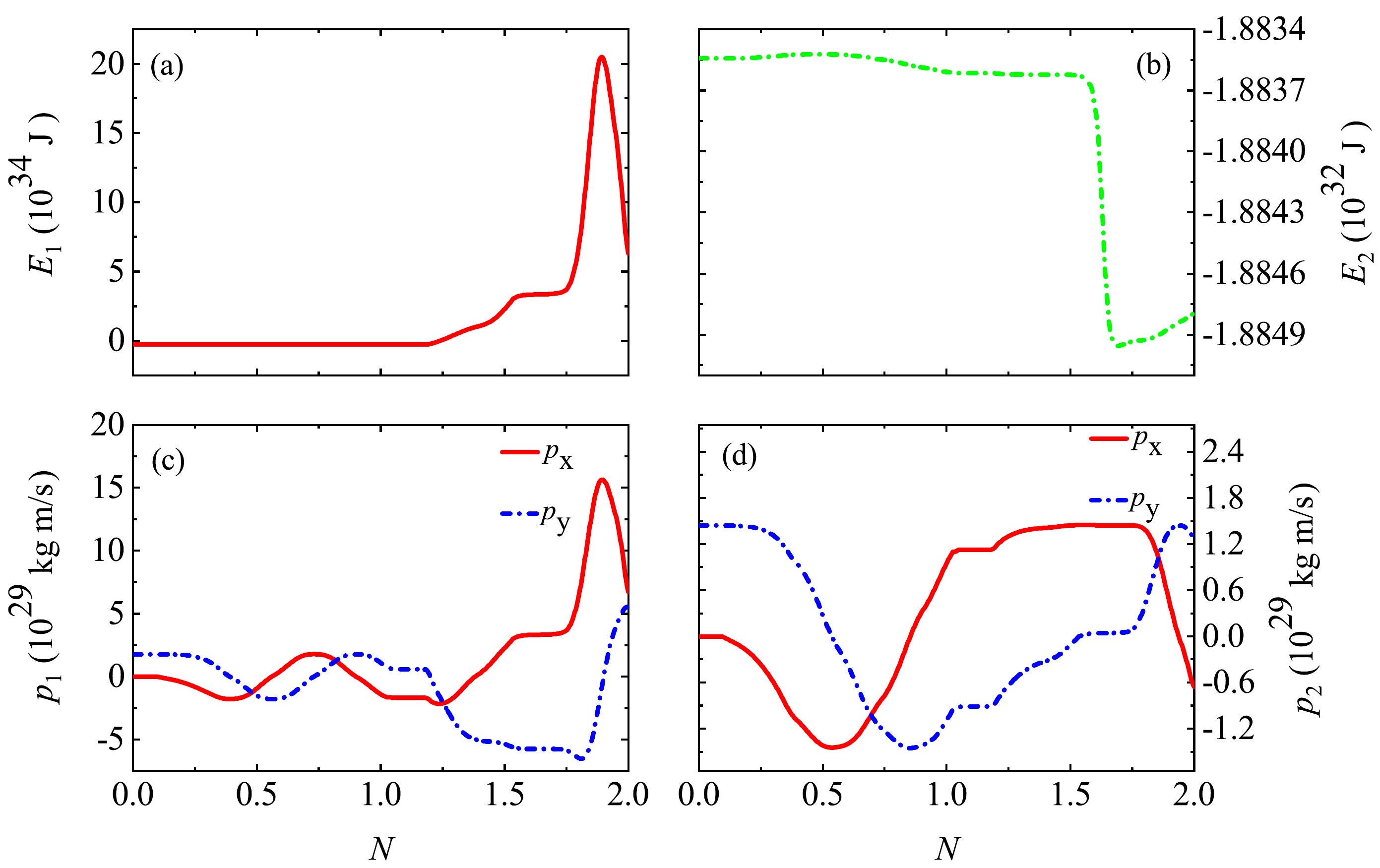}
\caption{\textbf{Attractive interacting case:}
energy of the (a) inner orbiting body and (b) outer orbiting body plotted vs number of Earth years ($N$). Two components of the two-dimensional momentum vector
(red solid line for the $x$ component and blue dash-dot line
for the $y$ component)
vs time counted in terms of the number of the Earth years
for (c)~the inner orbiting body
and (d)~the outer orbiting body.}
\label{fig:attenemom}
\end{figure*}

Due to the planetary collision,
Earth accelerates by following the non-orbiting path shown in Fig.~\ref{fig:attracdynamics}(d).
We choose to switch off the external thrust  after the collision,
which takes place 1.8 years after the start of the simulation, i.e., 0.6 years after the commencement of the thrust.
Subsequent to the collision,
we simulate the case of the thrust continuing to show clearly Earth's slow deceleration,
which appears in Fig.~\ref{fig:attenemom}(a).

We calculate the energy (discussed in Sec.~\ref{sec:two-orbiting-bodies}) of each of the two orbiting bodies and present the energy of each of the bodies in Figs.~\ref{fig:attenemom}(a) and (b), respectively, over two Earth years.
Initially, Earth's energy, 
presented in Fig.~\ref{fig:attenemom}(a),
depicts an almost-straight line displaying constant orbit energy due to the external thrust force being off during this time.
After the thrust is applied at 1.2 years,
energy increases as the external thrust force launches Earth towards the Sun,
where gravity assist propels the Earth to collide with Mars.
The energy peak in Fig.~\ref{fig:attenemom}(a)
indicates Earth's energy during its collision with Mars causing the Earth to accelerate even further from the Sun,
helped by continuation of the thrust force.
Figure~\ref{fig:attenemom}(b) shows the energy of Mars, where the steep falling line depicts the influence of Mars's strong attraction with Earth when Earth is nearby.

Now we present the momentum of each orbiting body when both bodies interact attractively. These time-dependent momenta are depicted in Figs.~\ref{fig:attenemom}(c) and~\ref{fig:attenemom}(d)
for the inner and outer bodies, respectively.
The $x$ and $y$ momenta are shown for each body as a function of the number of Earth years.
Earth momentum loci Fig.~\ref{fig:attenemom}(c) reveals that when the external thrust force is turned off, the Earth usually orbits the central body for the 1.2 years. After 1.2 Earth years, the external thrust force is applied to Earth which is clearly evident from their momentum loci.
The $x$ component of momentum Fig.~\ref{fig:attenemom}(c) depicts an almost-straight line that represents a strong attraction between Earth and Mars before they collide.
Figure~\ref{fig:attenemom}(d) shows the $x$ and $y$ momenta of Mars which represent Mars motion in the orbit.

In this subsection by analyzing planetary two orbiting bodies Earth and Mars, we explained the specific application of our attractive interaction framework. The external thrust force is applied to the inner-orbiting body and shows the dynamics before and beyond the collision. We study the momenta and energy of each orbiting body and identify the features in the subsequent plots.
The next subsection follows the same structure
of this subsection and the previous subsection.
\subsection{Noninteracting orbiting bodies}
In this subsection, we treat the case of two noninteracting orbiting bodies.
Specifically,
we reconsider the case of interacting bodies,
in our case Earth and Mars,
but neglect the attractive interplanetary interaction as a way to test our technique and intuition.
In this case,
the two orbiting bodies interact only with the central body that creates an attractive potential but the two orbiting do not interact with each other in the mathematical simulation.
This case is quite artificial but elucidates how attractive and repulsive forces meet in our formalism that accommodates both these cases.
In this case, we create controlled collision between planetary bodies by applying a time-dependent external force only to the lower orbiting body.
\begin{figure*}
 	\includegraphics[width=0.8\linewidth]{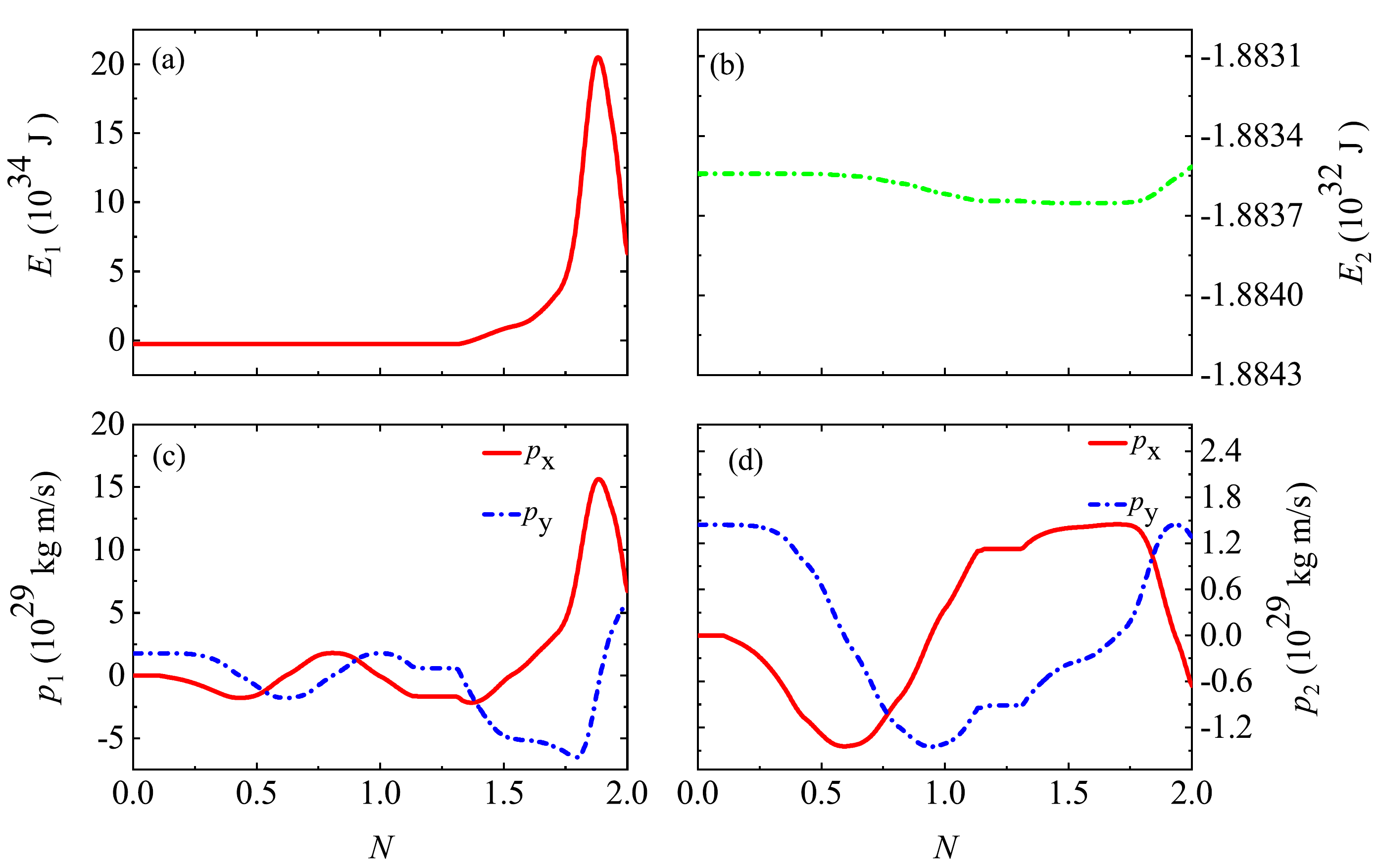}
\caption{\textbf{Noninteracting case:}
energy of the (a) inner orbiting body and (b) outer orbiting body plotted vs the number of Earth years ($N$).
Two components of the two-dimensional momentum vector
(red solid line for the $x$ component and blue dash-dot line
for the $y$ component)
vs time counted in terms of the number of Earth years
for (c)~the inner orbiting body
and (d)~the outer orbiting body.}
	\label{fig:nonattenemom}
\end{figure*}

We present our results for two noninteracting orbiting bodies 
with only the inner orbiting body being subjected to an external thrust force.
Concentric orbits with different eccentricities (0.017 for Earth and 0.093 for Mars) are considered but with the Earth-Mars interaction neglected.
Similar to the previous subsection,
Earth completes one cycle;
the external thrust force is opportunistically applied at 1.2 Earth years after commencement of the simulation.
The angular frequency,
strength of the thrust force,
and thrust-force function shape
are the same as in the previous subsection.
We ignore the mutually attractive force between Earth and Mars,
which is approximately $0.0099$\% of the total force produced by the interaction between the central body and each orbiting body.

Now we present the dynamics of the two noninteracting orbiting bodies.
As in the previous subsection, the external thrust force is first switched off for 1.2 Earth years
and turned on at this time when Mars is on the opposite side.
After applying the external thrust force, the Earth is first pushed closer to the Sun,
and the external thrust force plus the Sun's gravity assist launches Earth away from the Sun where it collides with Mars, a dynamic similar to that shown in Fig.~\ref{fig:attracdynamics}(c).

Now we calculate the energy (discussed in Sec.~\ref{sec:two-orbiting-bodies}) of two noninteracting orbiting bodies. Figures~\ref{fig:nonattenemom}(a) and~\ref{fig:nonattenemom}(b) show the energy of Earth and Mars, respectively, over two Earth years.
As in the previous subsection, Earth's energy is presented in Fig.~\ref{fig:nonattenemom}(a)
and displays an almost-straight line displaying constant orbit energy due to external thrust force being off during this time and then increases when the external thrust force launches Earth towards the Sun. 
The  energy in the plot peak of Fig.~\ref{fig:nonattenemom}(a) depicts the Earth's energy during its collision with Mars causing the Earth to accelerate even further from the Sun, helped by continuation of the thrust force.
Figure~\ref{fig:attenemom}(b) shows the energy of Mars, where in this case the steep falling line disappears, describing no attraction when Earth is nearby.

We present the results for our case of two noninteracting orbiting bodies
whose momenta are plotted over two Earth years.
The $x$ and $y$ momenta are shown for each body as a function of time quantified by Earth years.
Figure~\ref{fig:nonattenemom}(c) depicts the Earth momentum loci,
which shows that Earth moves in a normal orbit for 1.2 Earth years, but,
when the external thrust force is applied, Earth leaves the orbit.
As in the previous subsection, the $x$component of Earth's momentum did not depict a strong attraction with Mars before the collision.  
Mars's $x$- and $y$-momentum loci in Fig.~\ref{fig:nonattenemom}(d) shows that it moves independently in its orbit.  

In this subsection we have explained a specific application of our non-interactive framework by analyzing two planetary orbiting bodies Earth and Mars. The external thrust force is applied to the inner-orbiting body and we show the dynamics before and after collision 
while maintaining the thrust force well beyond the collision time to be able to check easily conservation of momentum and energy.
We study the momenta and energy of each orbiting body and describe features in the resultant plots.
\subsection{Verifying conservation laws}
In this subsection, we validate our results by using four different numerical methods;
validation holds if these four numerical results demonstrate sufficient convergence.
Conserved quantities such as energy and momentum are calculated by each numerical method and we present these results in a table and explain.
Specifically,
we validate the repulsive and attractive cases;
numerical results for the noninteracting case are quite similar to the attractive case so we do not need to validate this third case as validating the first two suffices.

Table~\ref{table:ta} shows the discrepancy for energy and momentum conservation in each of the repulsive (atomic) and attractive (planetary) cases as fractional errors,
i.e., how much the conservation rule has been violated,
which should only be within numerical approximation errors.
These four calculations are performed using four  numerical solvers in MATLAB\textsuperscript{\textregistered}.
Our calculations follow from Eqs.~(\ref{eq:feE}) and (\ref{eq:fep}) to obtain the fractional error in energy and momentum.
The first column in Table~\ref{table:ta} shows the four different numerical solver and we employ MATLAB\textsuperscript{\textregistered} terminology for these solvers.
\setlength{\tabcolsep}{9.5pt}
\renewcommand{\arraystretch}{1.3}
\begin{table}
\begin{tabular}{|c||c|c|c|c|}
\hline
\multirow{2}{*}{Solver}& \multicolumn{2}{|c|}{Repulsive} & \multicolumn{2}{|c|}{Attractive}\\\cline{2-5}
 {}&fe$(p)$&fe$(E)$&fe$(p)$&fe$(E)$\\
\hline
ode45&0.017802&0.71180&0.625&0.910\\
\hline
ode23&0.017803&0.71196&0.638&0.917\\
\hline
ode113&0.017804&0.71181&0.654&0.925\\
\hline
ode15s&0.017804&0.71183&0.654&0.925\\
\hline
\end{tabular}
\caption{%
Fractional error (fe) for momentum~$p$
and energy~$E$ for repulsive and attractive cases.
The first column shows numerical methods
using MATLAB\textsuperscript{\textregistered} terminology.
The second and third columns show fe for momentum
$p$ and energy~$E$ conservation in the repulsive case and the fourth and fifth columns for the attractive case.
The ODE MATLAB\textsuperscript{\textregistered} solvers are ode45 for the explicit Runge-Kutta (4,5) method,
ode23 for the explicit Runge-Kutta (2,3) method,
ode113 for the  variable-step, variable-order Adams-Bashforth-Moulton PECE solver
and ode15s for the backward differentiation formula
(also known as Gear's method).} 
\label{table:ta}
\end{table}

These results in Table~\ref{table:ta} closely agree for all four numerical methods applied to the repulsive and attractive cases for two orbiting bodies with ode45 yielding the best result in each case.
Thus
our numerical results validate our ode45 solver results.

\section{Discussion}
\label{Sec:5}
We have developed an approach to create a deterministic collision between two orbiting bodies by considering three cases:
where the two bodies are mutually attractive or repulsive or noninteracting.
Furthermore we consider, for the repulsive case,
both instances of a force acting only on the lower body,
which pertains to each of the three cases,
and to the force applying equally to both bodies,
which is pertinent to the case of controlling two-electron atoms in a helium-like atom.

Our method involves writing coupled differential equations that include an external driving force with a convenient time-dependent shape.
The task of solving force parameters is achieved by using the Runge-Kutta (4,5) method for differential equations.
We validate these results using alternative numerical methods and thus show that we have a trustworthy, accurate method for causing collisions.
As examples,
we treat a classical two-electron helium atom and planetary collisions in our solar system.
Although our theory is based on standard principles of classical mechanics,
we have successfully addressed the challenges of devising one method for the distinct scenarios of both repulsive and attractive central potentials,
devised an efficient method for obtaining successful control sequences that work in both scenarios,
incorporated energy and momentum of the control pulses into the equations to ensure simulations meet stringent conservation laws,
and used these conservation laws to validate our results.

Typically the helium atom would be treated quantum mechanically,
but a classical treatment becomes increasingly valid for highly excited electrons.
In such cases,
a classical treatment could be valuable for applying control techniques to highly excited two-electron atoms,
which are known sometimes as two-electron Rydberg atoms.
Our choice of force pulse shape for causing electron-electron collisions is commensurate with past treatments of inducing collisions and so should be feasible.
As the size of the external force is much broader than the size of the atom in the repulsive case, we include the case where an external force is acting on both bodies.
As repulsive bodies can collide without genuine contact,
we have been careful to introduce a rigorous definition of collision.

In the attractive case,
we discuss how a controlled thrust could launch Earth into a collision with Mars.
Although this scenario is infeasible and undesirable,
the didactic appeal of relating our method to tangibility of our planetary motion makes our approach intuitively clear.
More practically,
our method relates to rocket launches,
and we have cited appropriate references for making this connection.
Notably,
our technique reveals a gravitational assist,
which arises directly and naturally from our numerics without having had to insert this method artificially.

As an artificial but instructive case,
we study the Earth-Mars collision while ignoring their mutual gravitational attraction.
In this noninteracting case,
we obtain almost identical results for thrust force and motion except that the attractive case shows Earth and Mars would slow down during the collision and then move apart whereas,
in the noninteracting case,
Earth and Mars do not slow down during the collision.
\section{Conclusion}
\label{Sec:6}
In conclusion,
we have introduced a unified mathematical framework for solving two orbiting bodies in a central potential for three different cases, i.e.,  repulsive, attractive and noninteracting orbiting bodies,
with an external force applied to one or both bodies.
This framework is used to solve driving-force parameters to cause deterministic collisions,
and we validate our results by showing that conservation laws hold up to numerical error.
Furthermore,
we have been careful to define collisions mathematically so that contact is not required and then show these collisions in momentum and energy plots.

Our approach to electron-electron collisions,
pertinent to the case that both electrons are ``planetary,"
is a radical departure from standard electron-electron collision theory and could open new avenues for studying electronic properties of effective two-electron atoms.
Although planetary-atom behavior in helium-like atoms has been studied in some ways,
collisions between planetary electrons in helium have not been studied before,
and our classical model provides insight and guidance on how to create deterministic collisions,
but of course a fully quantum treatment is needed to obtain exquisitely accurate laser-pulse design for quantum control.
Embarking into collision dynamics between planetary electrons of helium
could help to elucidate and quantify the electronic structure and usher in new atomic phenomenology. 
Our classical analysis is a zeroth-order approximation to a full quantum treatment and inspires the possibility of using Perelomov coherent states~\cite{perelomov1986,perelomov1977}
to obtain more accurate pulse design by perturbing around coherent states that closely approximate mean dynamics of planetary electrons over several orbits.
In contrast to previous work,
which can involve ensembles of classical trajectories or, alternatively,
wave-packet dynamics in a quantum treatment,
our approach is strictly classical and deterministic with a single pair of orbiting bodies.
The gravitational case of Earth-Mars collisions is fanciful but didactically valuable.
Our approach can serve as a foundation for practical controlled collisions between orbiting bodies.
\section*{Acknowledgements}
AM acknowledges the Chinese Scholarship Council
(CSC) for financial support.
BCS appreciates financial support from the National Natural Science Foundation of China (NSFC) (Grant No. 11675164).
\bibliographystyle{apsrev4-2}
%

\end{document}